# Undercooling growth and magnetic characterization of ferromagnetic shape memory alloy $Ni_2FeGa$ single crystals

J. F. Qian, H. G. Zhang, J. L. Chen, W. H. Wang and G. H. Wu[1]

Beijing National Laboratory for Condensed Matter Physics, Institute of Physics, Chinese Academy of Sciences, Beijing 100190, People's Republic of China

**Abstract**

$Ni_2FeGa$ single crystals have been grown in undercooling conditions provided by a glass-purification method. It has been found that trace amounts of γ phase embededin the single crystalline matrix preferentially orients in the <100> orientation along the growth direction. This γ phase generates directional residual stress and results in an anisotropic two-way shape memory effect. Large strains of -2.5% in the [001] and 1.5% in the [010] directions have been observed. This trace γ phase also improves the ductility of the material and thus the crystals could be plastically deformed at room temperature in the parent phase. The <110> and <111> orientations in $Ni_2FeGa$ alloy were identified as the easy and hard magnetization directions, respectively, in the parent phase by using low field M-T measurements.

## 1. Introduction

---

In the last decade, ferromagnetic shape memory alloys (FSMAs) have attracted considerable attention because of their potential applications as magnetic actuator devices. Up to now, several candidates for FSMAs have been exploited including Fe-Pd [1], Fe-Pt [2], and some Heusler alloys (with a stoichiometry formula of $X_2YZ$ where X and Y are the transition metals and Z is a main group element), such as $Ni_2MnGa(Al)$ [3-6], $Co_2NiGa$ [7], Fe-Mn-Ga [8] and $Ni_2Co(Fe)Ga$ [9]. Among them, $Ni_2FeGa$ and its off-stoichiometry alloys are very promising materials due to their appreciable ductility [10-14]. Single crystal samples are ideal objects for the investigation of fundamental properties, such as martensitic transformation, magnetic anisotropy and mechanical deformation. However, the stoichiometric $Ni_2FeGa$ single crystal is difficult to fabricate, because its solidification usually causes the appearance of a second phase, the γ phase, with a face centered cubic (fcc) structure [13,15,16]. The formation of the γ phase will seriously disturb the normal crystal growth. Therefore, the NiFeGa alloys prepared in most previous works were with off-stoichiometry compositions [11,17-19].

In our previous work, the results indicated that a fast cooling condition can mostly eliminate the appearance of the γ phase in polycrystalline $Ni_2FeGa$ alloys [13]. This is similar to the solidification behavior in other systems where the peritectic reaction occurred [9,20]. However, the single crystal growth of $Ni_2FeGa$ in an undercooling condition has not been investigated systemically and thus the question of how to eliminate the γ phase in single crystal $Ni_2FeGa$ is still open.

In this letter, we report the growth of stoichiometric $Ni_2FeGa$ single crystals in an undercooling condition provided by the glass-purifying method. By using this method, the γ phase can be mostly eliminated, similar to the case in which the samples are prepared by a melt-spinning method [21]. The single crystals show a large anisotropic strain in a two-way shape memory effect and an improved ductility, which can be attributed to the trace amount of preferentially oriented γ phase remaining in the crystals. Based on this kind of microstructure, the magnetic anisotropy of the alloys has also been investigated.

2. **Experiment**

About 30 g of starting materials with the composition of $Ni_2FeGa$ were prepared by arc-melting Ni, Fe, and Ga metals (the purity is higher than 99.95%). Crystals were grown by Czochralski method in the MCGS-3 instrument with a magnetic levitation cold-crucible system and high frequency induction-heating unit. Single crystal bars 3×3×10 (length) $mm^3$ with the various orientations, <001>, <111>, and <110> were used as seeds. The growth rates of 15–25 mm/h and the rotation rate of 30 rpm were adopted. The crystal structures of the bulk and powder samples were determined by X-ray diffraction using Cu-Ka radiation at room temperature. Laue diffractograms were obtained by the x-ray back-reflection technique from different positions across one transverse section of the single crystals to align their orientation. Metallographic observation was performed on the crystal surface which was electrochemically polished by 20% $HNO_3$ alcohol solution to inspect the quality and the phase purity of the grown crystals. The magnetic properties were measured by using a superconducting quantum interference device (SQUID) magnetometer

(Quantum Design). The transformation strain in different orientations was measured by a strain gauge with the standard Ni coil.

In the present work, the single crystals were grown in an undercooling condition provided by the glass-purification method. During growth, the melt was immersed in molten glass to purify the impurities that usually promote the solidifying nucleation. This will lower the solidification point of the melt and lead to the undercooling condition[22].

The glass-purifier, consisting of 50% $B_2O_3$, 30% $Na_2SiO_3$ and 20% $Na_2B4O_7$, was made from high purity and dehydrated reagents by keeping them at a molten state at 900 $^oC$ for 24 h. A quartz bowl was used to contain the starting material and the glass, which was settled in the cold-crucible for growth. The starting materials were melted in the crucible and remained there for about a half-hour before growth. The growth chamber of the instrument was pre-vacuumed to the level of about $3x10^{-4}$ Pa before heating the starting material, and an argon atmosphere with a partial pressure slightly higher than the environment atmosphere was maintained during the growth of $Ni_2FeGa$ crystals.

For comparison, the undercooling solidification with a glass-purifier and the growth without a glass-purifier were performed in parallel in the same equipment. In the undercooling solidification, the liquid was superheated to a temperature of about 200-300 $^oC$ higher than the melting point and held at this temperature for about 2 minutes. After superheating, the melts were cooled at a rate of 50$^o$C/min. until the recoalescence was observed. This cycle between superheating and undercooling was performed three times. The melt temperature was measured using an infrared temperature sensing system. The various thermal parameters were identified by averaging the data measured from several parallel experiments. The chemical analysis was carried out to evaluate the composition change in the grown crystals and residual material. The as-grown crystals were annealed at 577 $^{o}$C for four days and then quenched into ice water to obtain a high-degree of atomic ordering.

3. **Results and discussion**

As shown in Figure 1(a), Ni$_2$FeGa single crystals were obtained with diameters of 6–12 mm and a maximum length of 25 mm. The single crystal rod clearly exhibits flat growth interfaces and symmetric facets, indicating a good growth condition. Figure 1(b) shows a typical Laue diffractogram in <110> orientation; no sign of polycrystallines or a second phase can be observed. Chemical analysis indicates that there is no significant change of chemical composition in the grown single crystals or the residual materials; the composition deviation in all samples is less than 2.0 at. %. These results indicate that the growth of single crystal Ni$_2$FeGa has been successfully realized by using this undercooling method. A schematic of the melt immersed in liquid glass is presented in Figure 1(c).

In the present work, Ni$_2$FeGa rods were also grown using the same cold-crucible without a glass-purifier. As shown in Figure 2(a), the well-developed second phase can be observed by the metallographic microscopy on the etched surface of the sample. This result confirms that the observed second phase originates from the solidification. Such second phase leads the grown Ni$_2$FeGa rod to be polycrystalline.

On the other hand, no grain boundary can be observed in the single crystals grown in the undercooling condition, while only very tiny etch pits with a density of about $10^6$ cm$^{-2}$ appeared, as shown in Figure 2(b).

In the ternary Ni-Fe-Ga system, the equilibrium phase diagrams show that surrounding the stoichiometrical composition of Ni$_2$FeGa, there is quite a large composition range with mixed γ and fcc phases, reaching temperatures of at least 1100 °C[16]. Our previous work indicted that the γ phase can be eliminated from Ni$_2$FeGa samples by using melt-spinning method[13], which was attributed to the undercooling condition provided by the fast solidification during the process. Researchers have previously used the undercooling method to eliminate the γ phase in some similar systems, like Co$_2$NiGa [20] and NiCoFeGa[9] and indicated that solidification behavior belongs to peritectic reaction. Therefore, it is reasonable to speculate that the solidification behavior in Ni$_2$FeGa system is also a peritectic reaction.

The undercooling solidification has been performed in the present work. The results indicate that the melting point of the Ni$_2$FeGa starting material is of about 1252 °C and the recoalescence temperature is about 1070 °C, showing an undercooling of about 180 °C. The results of metallographic observation on these samples indicate that the phase situation is similar to that shown in Fig. 2(b), which shows the γ phase has been mostly eliminated by the undercooling condition provided by the glass-purification method. Using a temperature sensing system, the growth temperature was also measured. The temperature at the surface of the melt is about

1180 °C. The temperature at the solid-liquid interface should be lower than this value because of the heat dissipation from the crystal growth and the seed holder. These observations strongly suggest that our $Ni_2FeGa$ single crystals are grown in a dynamic undercooling state. In this system, about 70 °C of undercooling is enough to suppress the nucleation of the second phase. Usually, a second phase can be eliminated by the undercooling condition, if the peritectic reaction is not very serious.

Figures 3(a) and (b) show the powder XRD results for the $Ni_2FeGa$ samples grown without and with the glass-purifier, respectively. The XRD pattern of the sample grown without the glass-purifier shows clearly two sets of diffraction peaks. In indexing these peaks, the coexistence of a body centered cubic (bcc) phase and fcc phase is confirmed [23]. This further confirms that the second phase observed in Figure 2(a) is the $\gamma$ phase. On the other hand, only the bcc phase can be observed in the XRD pattern of the single crystal sample grown in an undercooling condition with the glass-purifier, as shown in Fig. 3(b).

However, further examination of the single crystal samples by using an XRD step-scan technique reveals that trace amounts of the $\gamma$ phase was still observed on the (110) etched surface. As shown in Figure 4, a very weak [200] peak of the $\gamma$ phase, neighboring the strong [220] peak of bcc phase, was observed on the (110) surface of the single crystal. This result implies that the etch pits shown in Figure 2(b) are derived from the $\gamma$ phase, where high stress causes an acceleration of the chemical etching. It should be emphasized that the (110) surface used for the step-scan XRD is the grown surface, namely, this single crystal sample was grown along <110>

orientation. Therefore, we conclude that the γ phase, with quite a small size and low density still exists, embedded in our single crystals and preferentially orients along the growth direction of the single crystal, i.e. <100> orientation. This phenomenon has also been observed in other as-grown single crystals with various orientations, which indicates that the oriented γ phase is generated in a local peritectic environment on the solid-liquid boundary caused by the temperature-composition fluctuation during the growth, rather than random precipitation during the heat treatment. Interestingly, we found that <100> direction is always the preferential orientation of the γ phase, independent of the growth direction of single crystal.

Figure 5 shows the temperature dependences of magnetization of $Ni_2FeGa$ single crystal samples oriented in different directions (<111>, <110> and <100>). The curves were measured in the field of 50, 500 and 50 kOe, respectively. The saturation magnetization of the martensite in $Ni_2FeGa$ is slightly higher than that of the parent phase (see the inset). Under low field, the style of the M-T curves is dominated by the difference of magnetocrystalline anisotropy between the parent and martensite phases [11]. The dramatical decrease of the magnetization at 95 K indicates the occurrence of the martensitic transformation, which has a thermal hysteresis of about 10 K. From the curves measured at 50 and 500 Oe (see Figure 5), one may find that the martensite in the <110> orientation has lower anisotropy than that of the other two orientations in this field range. For the parent phase, the curves measured at 500 Oe show that the <110> is the easy direction of magnetization, while <111> is the hard one. However, this anisotropic property cannot be identified in the curves measured at 50 Oe, due to the similar barrier energies of magnetic domain walls and the Zeeman energy[4].

Since the various variants oriented in the different directions, the martensite shows a complex anisotropic property. As shown in Figure 6(a), the <110> direction has low anisotropy under 2 kOe but becomes a relatively hard direction of magnetization at higher fields. This transition of anisotropy may be attributed to the restriction of martensitic variant to magnetic domain in the various orientations[24]. In contrast, it is quite difficult to identify the anisotropy among the three orientations from the M-H curves of the parent phase, as shown in Figure 6(b). This is due to its cubic structure with high symmetry.

A two-way shape memory effect has been observed in our $Ni_2FeGa$ samples of single crystals. As shown in Figure 7, the <001> orientation (growth direction) shows a contraction of 2.5% through the martensitic transformation, which is completely recovered during the reverse transformation. This magnitude of deformation is much larger than that of polycrystalline samples [13]. In the <010> orientation, which is perpendicular to [001] direction, the 1.5% strain is positive, indicating an expansion occurred during the martensitic transformation. This orientation-dependent behavior has also been observed in single crystal $Ni_2MnGa$, where it was attributed to the residual stress induced by the defects included in the crystals [5]. In our $Ni_2FeGa$ single crystals, the <100> preferentially oriented γ phase plays an important role in generating a directional residual stress and thus leads to an anisotropic strain behavior. In <110> orientation, a positive strain of about 1.2% has also been obtained.

Due to the existence of the oriented γ phase in the $Ni_2FeGa$ single crystal, we expect that the ductility of this material can be enhanced. Figures 8 (a)-(c) show the visual evidence of the shape memory effect in a pre-deformed $Ni_2FeGa$ sample. As shown in Fig. 8(a), a <110> oriented single crystal bar was first bent severely at room temperature, where the sample is still in the parent phase because the transformation temperature is too low for a superelastic effect to occur. Then the bent sample was straightened after immersion in liquid nitrogen, where the martensitic transformation occurred, as shown in Fig. 8(b). Interestingly, as shown in Fig. 8(c), when the sample was removed from the liquid nitrogen, the straight sample recovered its original deformed shape while heating through the reverse transformation. This clearly reveals a typical shape memory effect. These results indicate that a trace of the γ phase could improve the ductility, without damaging the martensitic transformation of the $Ni_2FeGa$ alloy.

## 4. Conclusions

In summary, the stoichiometric $Ni_2FeGa$ single crystals have been grown in an undercooling condition by using a glass-purification method to eliminate the peritectic γ phase. The undercooling experiments confirm that the glass purifying provides an undercooling condition which is good for the pure bcc phase solidification and single crystal growth. We find that the oriented γ phase uniformly embeds in the matrix and preferentially orients along the <100> growth direction. Such trace amounts of the γ phase generate a directional residual stress which leads to an anisotropic strain behavior in the two-way shape memory. The existence of such a γ phase also enhances the ductility of the single crystals, beneficial for applications in actuators. For the single crystal samples, the <110> orientation of the parent phase is identified as the easy direction of magnetization based on low field M-T measurements. A complex anisotropy behavior has also been observed in the martensite of $Ni_2FeGa$, which reflects the interactions between martensitic variants and the magnetic domains.

**Acknowledgments**

This work is supported by the National Natural Science Foundation of China in Grant No. 51021061 and 11174352 and National Basic Research Program of China (973 Program, 2009CB929501).

Figure Captions

Figure 1. (a) As grown single crystal rod; (b) Laue diffractogram on (110) plane; and (c) and quartz bowl loaded with starting material and glass (after growth).

Figure 2. Optical micrographs of the $Ni_2FeGa$ samples grown without (a) and with (b) glass-purifier, respectively.

Figure 3. Powder XRD patterns of the samples grown by the same magnetic levitation cold crucible system without (a) and with (b) glass-purifier.

Figure 4. Step-scan XRD pattern for the bulk single crystal sample, inset shows the enlarged peak around 50° marked by the arrow.

Figure 5. Temperature dependence of magnetization of $Ni_2FeGa$ single crystal samples measured in the magnetic field of 50 Oe (a), 500 Oe (b) and 50 kOe (inset).

Figure 6. Magnetization behaviors of the martensite (a) and the parent phase (b) in three orientations measured in the low field range.

Figure 7. Temperature dependence of strain measured on the $Ni_2FeGa$ single crystalline samples oriented in [001], [010] and [110].

Figure 8. Visual evidence of shape memory effect occurred on the pre-deformed Ni$_2$FeGa single crystal samples. The martensitic transformation temperature (*Ms*) is 95 K, as well as the austenite transformation start (As) is 105 K.

**Figure 1**
**Click here to download high resolution image**

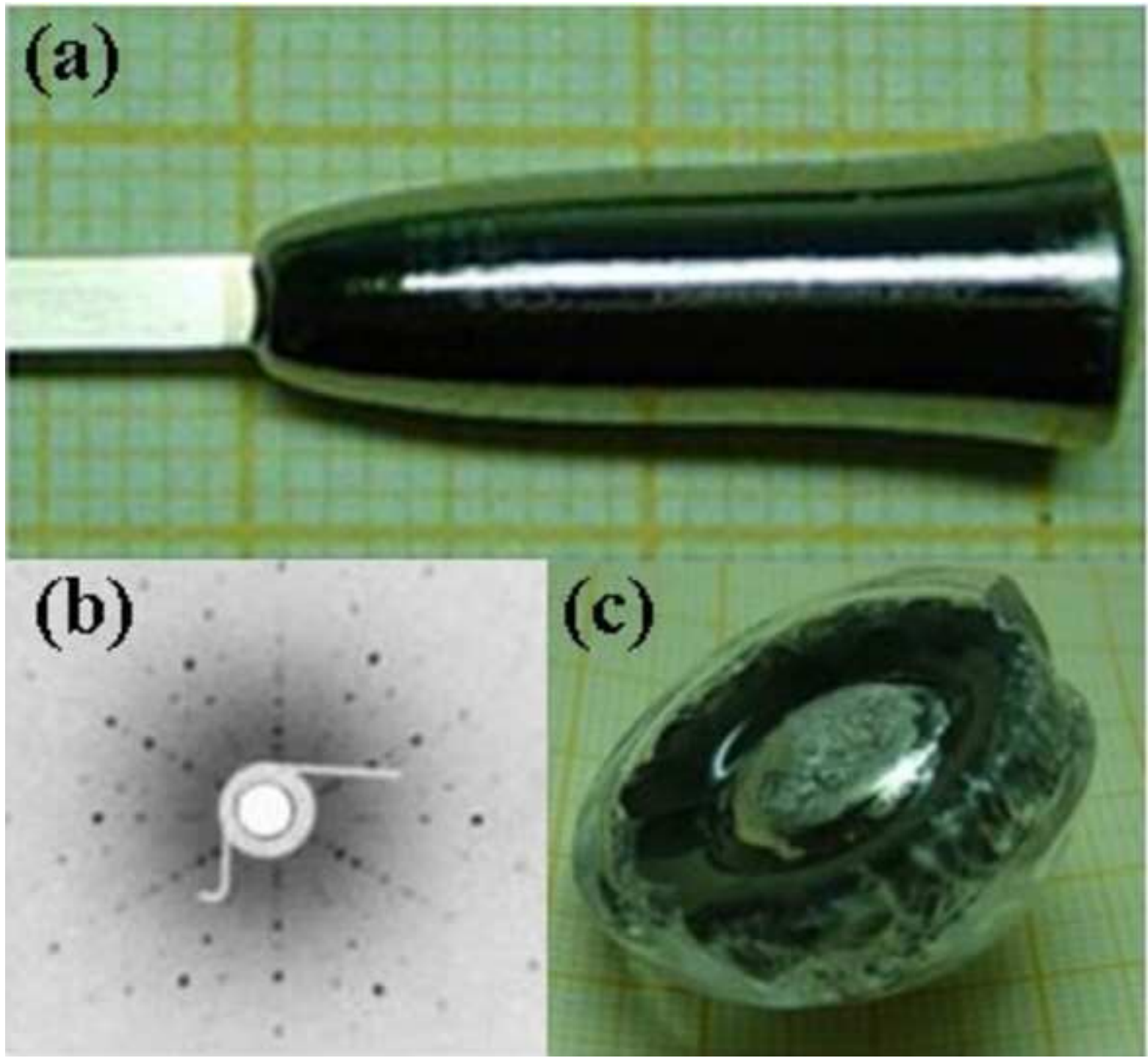

Figure 2
Click here to download high resolution image

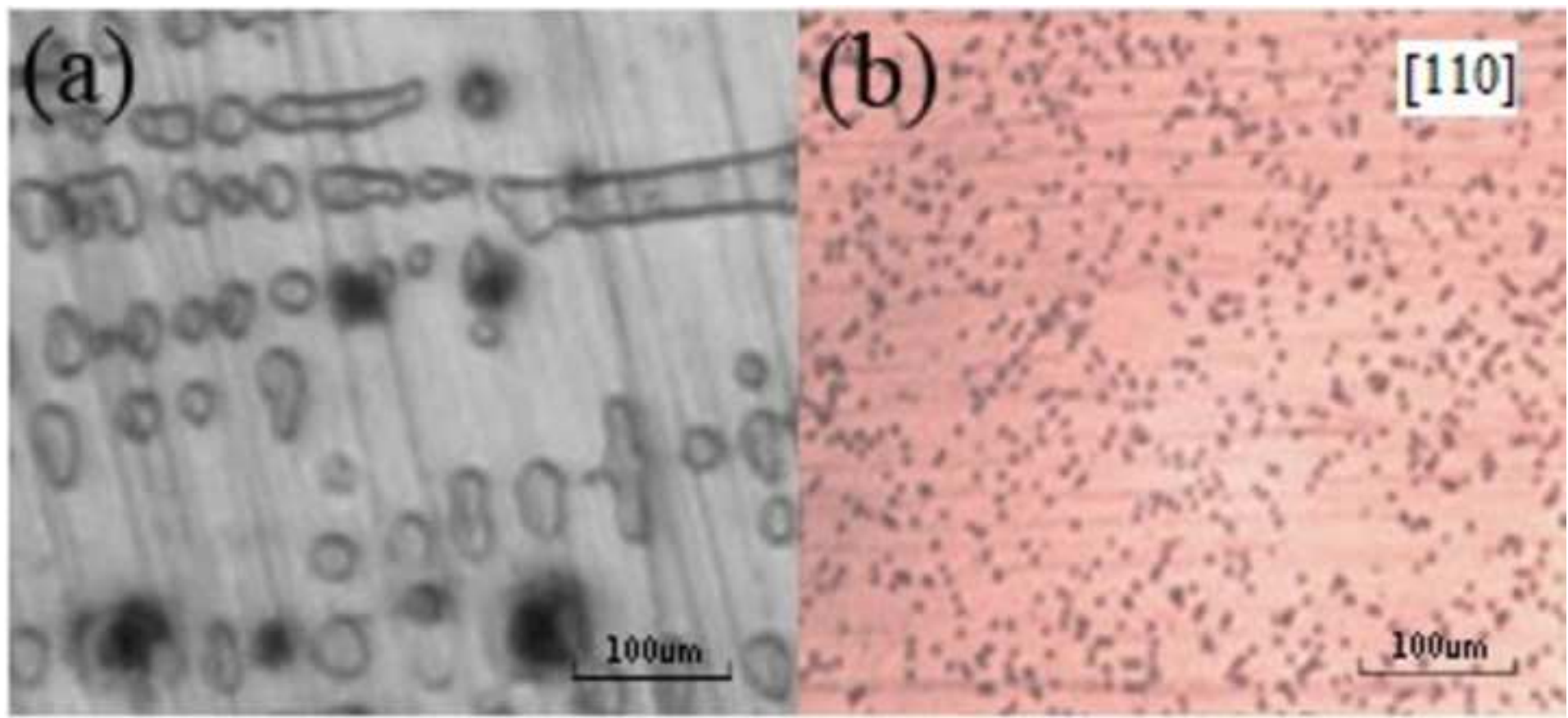

**Figure 3**
[Click here to download high resolution image](#)

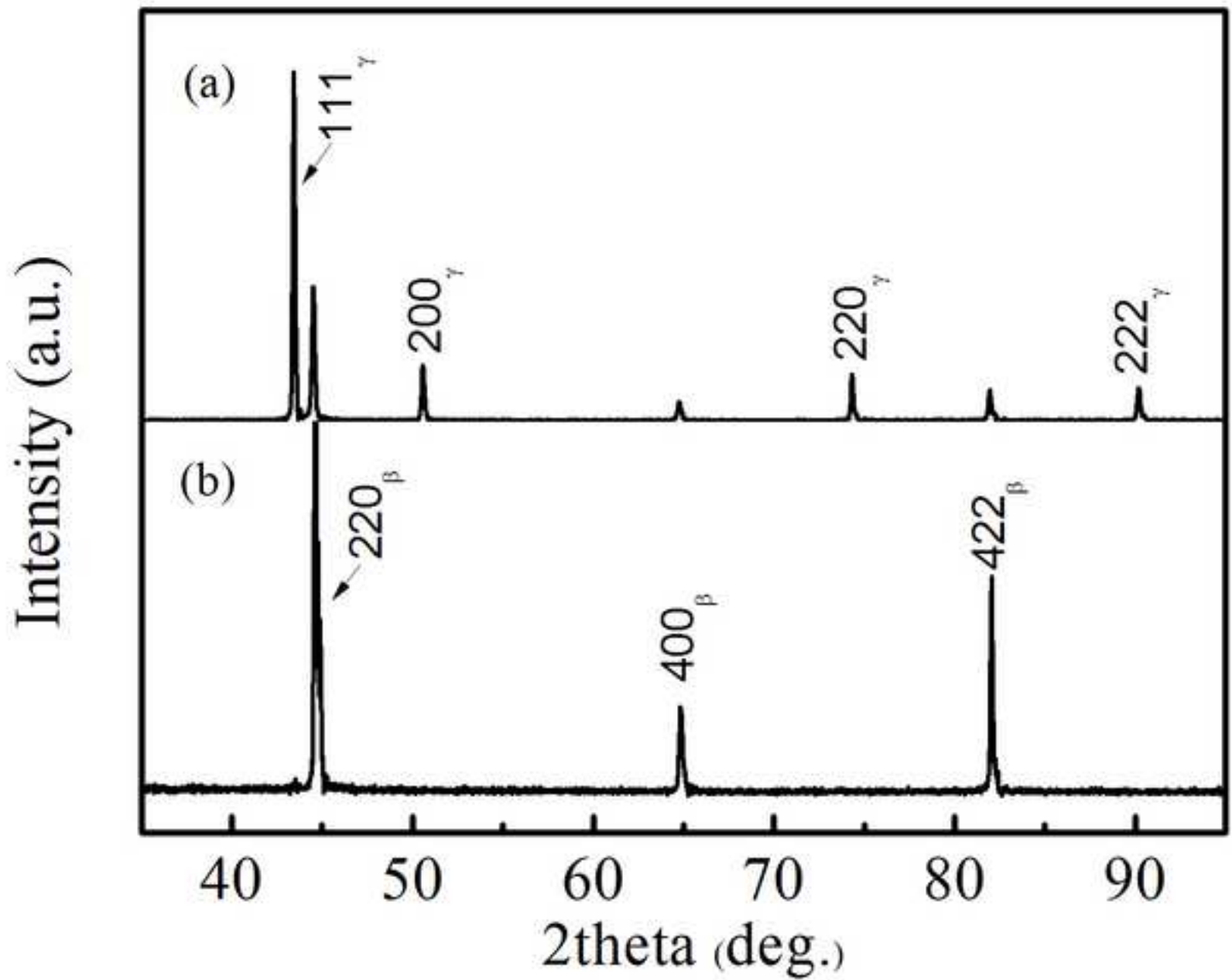

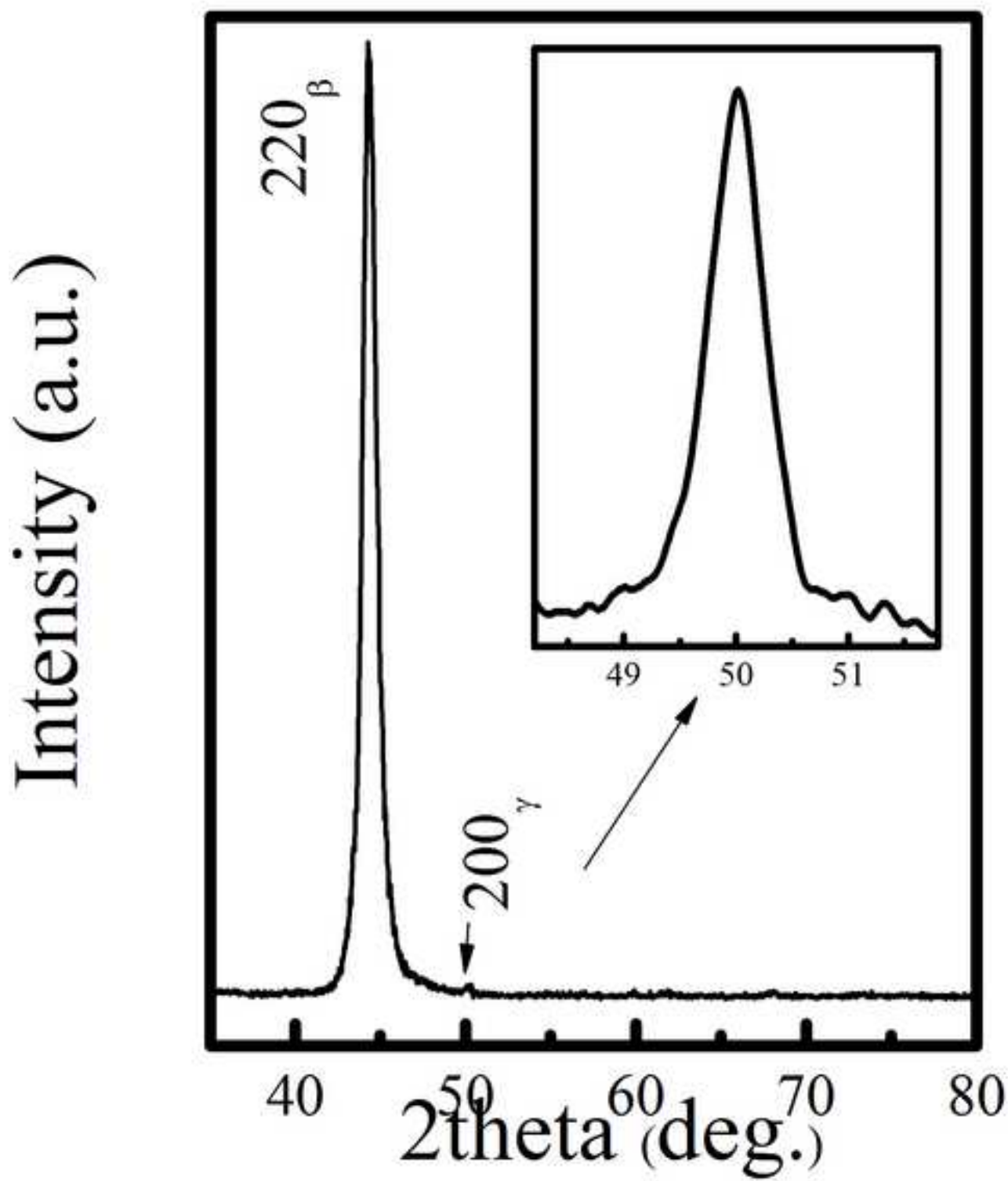

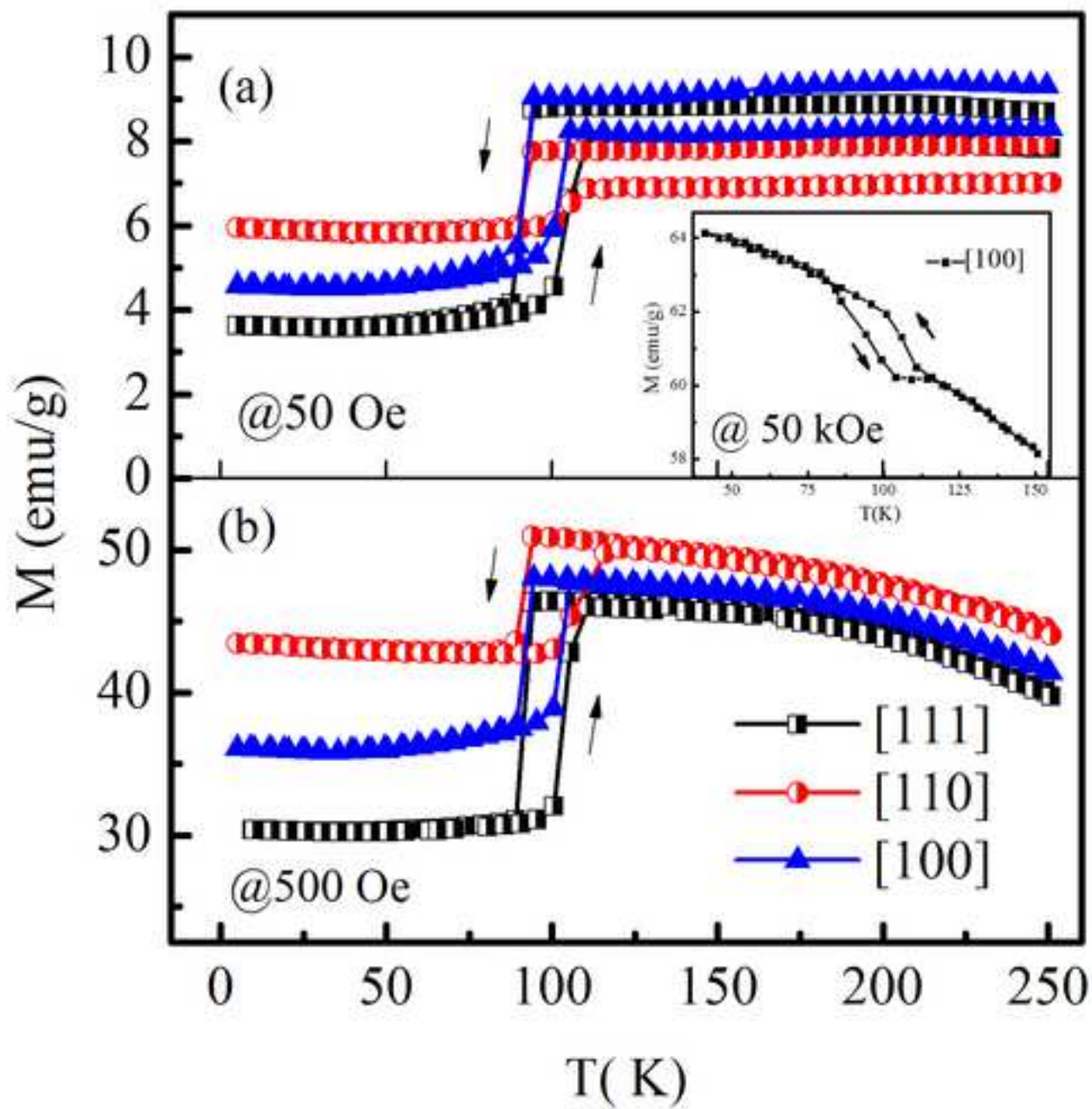

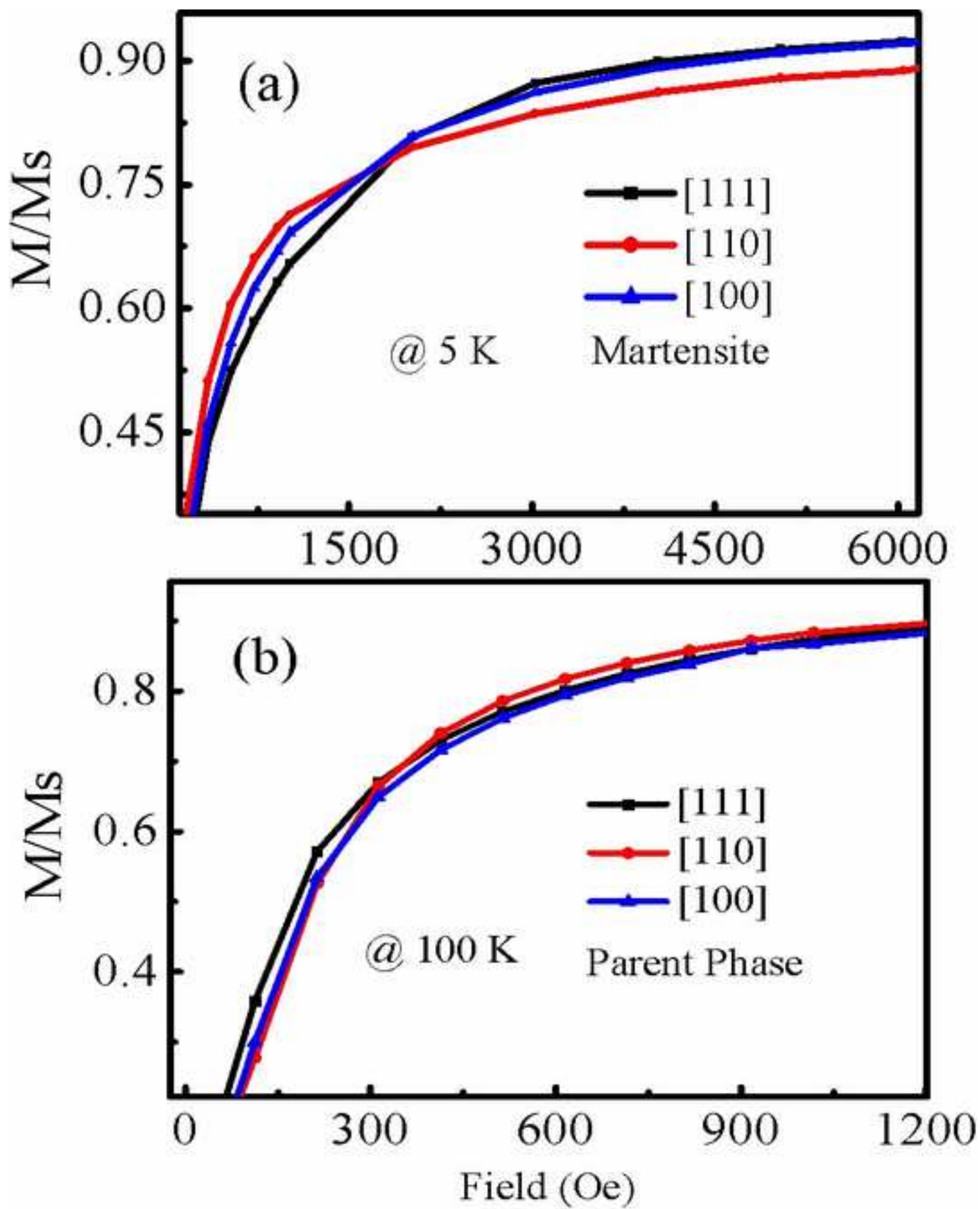

**Figure 7**
**Click here to download high resolution image**

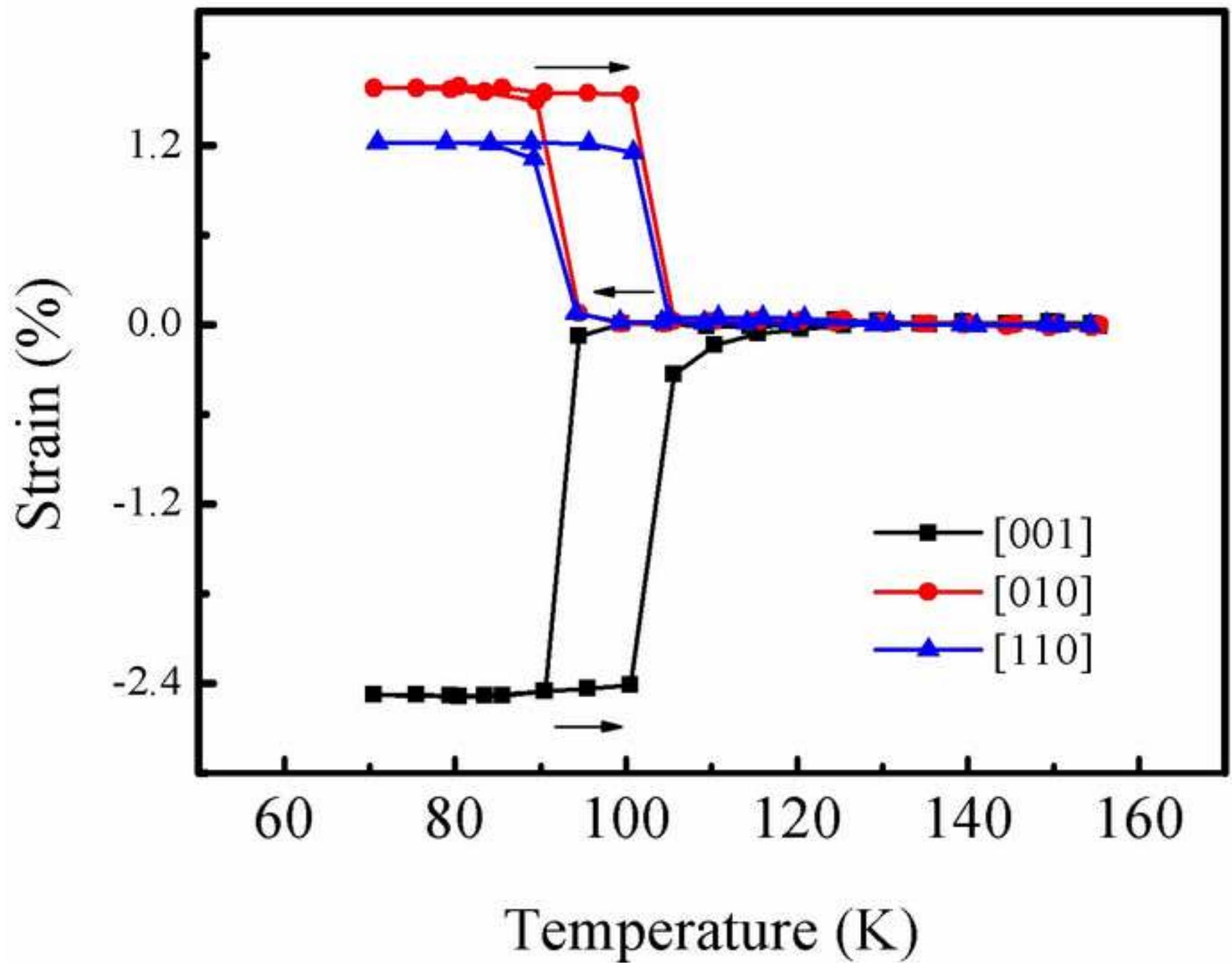

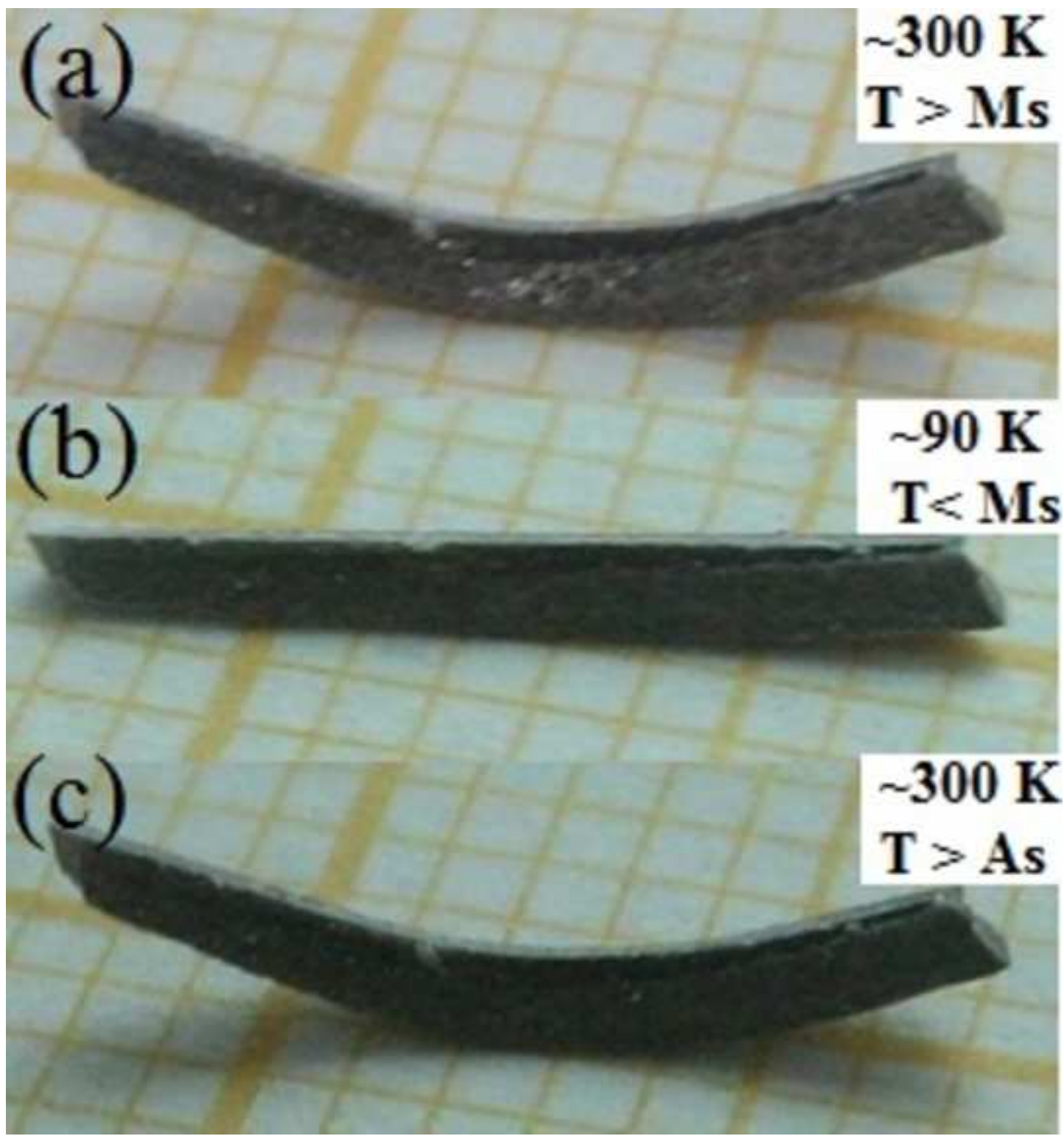